\documentstyle[preprint,prb,aps,epsf,eqsecnum]{revtex}
\begin{document}
\tightenlines
\title{Completing Bethe's equations at roots of unity}
\author{Klaus Fabricius
\footnote{e-mail Fabricius@theorie.physik.uni-wuppertal.de}}
\address{ Physics Department, University of Wuppertal, 
42097 Wuppertal, Germany}
\author{Barry~M.~McCoy
\footnote{e-mail mccoy@insti.physics.sunysb.edu}}               
\address{C.N. Yang Institute for Theoretical Physics, 
State University of New York,
 Stony Brook,  NY 11794-3840}
\date{\today}
\preprint{YITP-SB-00-89}
\maketitle

\begin{abstract}

In a previous paper we demonstrated that Bethe's equations are not
sufficient to specify the eigenvectors of the XXZ model at roots of
unity for states where the Hamiltonian has degenerate eigenvalues.
We here find the equations which will complete the specification of
the eigenvectors in these degenerate cases and present  
evidence that the $sl_2$ loop algebra symmetry is sufficiently powerful
to determine that the highest weight of each irreducible representation 
is given by Bethe's ansatz.  

\end{abstract}

\section{Introduction}

The XXZ model with periodic boundary conditions, defined by
\begin{equation}
H=-{1\over 2}\sum_{j=1}^L (\sigma^x_j\sigma^x_{j+1}+
\sigma^y_j\sigma^y_{j+1}-\cos\gamma \sigma^z_j\sigma^z_{j+1}),
\label{ham}
\end{equation}
where $\sigma_j^i$ is the $i$ Pauli spin matrix at site $j$ and
$j=L+1\equiv 1,$ 
has long been studied by means of what has come to be called Bethe's equation 
\begin{equation}
\left({\sinh {1\over 2}(v_j+i\gamma)\over \sinh{1\over 2}
(v_j-i \gamma)}\right)^L=\prod_{l=1\atop l\neq j}^{{L\over 2}-|S^z|}
{\sinh{1\over 2}(v_j-v_l+2i\gamma)\over\sinh{1\over 2}(v_j-v_l-2i\gamma)}
\label{beq}
\end{equation}
where $S^z={1\over 2}\sum_{j=1}^L\sigma^z_j$ is a conserved quantum
number.
The eigenvalues of (\ref{ham}) are given by
\begin{equation}
E={\cos\gamma\over 2}L-2\sum_{j=1}^{{L\over 2}-
|S^z|}{\sin^2\gamma\over \cosh v_j-\cos\gamma}
\label{energy}
\end{equation}
and the eigenvectors are given in terms of the $v_j.$

The authors who originally derived these equations
 \cite{bethe}-\cite{yy2}  concentrated on the energy and eigenvector of 
the ground state and this
computation has been done with great rigor and clarity. However, the
derivations seem to superficially extend to all eigenstates and for
almost three decades there have been extensive efforts
 \cite{des},\cite{tak2}-\cite{nlk}  made to study these 
excited energy states and for generic values of $\gamma$ it has been 
proven \cite{lanst1} - \cite{lanst2} that the solutions to (\ref{beq}) do
in fact give all the $2^L$ eigenvalues of (\ref{ham}).

On the other hand it was shown by Bethe \cite{bethe} in his original 
study of the case $\gamma=0$ that for the Heisenberg antiferromagnet the 
equation (\ref{beq}) with finite $v_j$  does not give all eigenvalues. 
Furthermore it was realized \cite{baxc} as early as 1973  
that the equation (\ref{beq}) will not specify solutions of the form
where both the numerator and denominator  simultaneously vanish and
that these special solutions do indeed occur in the ``root of unity case''
\begin{equation}
\gamma_0={r\over N}\pi
\label{root}
\end{equation}
where $r$ and $N$ are relatively prime and $1\leq r\leq N-1.$
Therefore there are indeed values of $\gamma$ which are not generic in the 
sense of \cite{lanst1}-\cite{lanst2} where Bethe's equation (\ref{beq}) 
with finite
$v_j$ is known not to give all eigenstates of the system.

Recently it was demonstrated \cite{dfm} that when the root of unity
condition (\ref{root}) holds that the Hamiltonian (\ref{ham}) has an
invariance under the $sl_2$ loop algebra and that this symmetry leads
to degenerate multiplets of energy eigenvalues. Moreover we
subsequently showed \cite{fm} when $\gamma\rightarrow \gamma_0$ that
there are solutions $v_k$ to (\ref{beq}) which have $N$ members of the
form
\begin{equation}
v_k=\alpha+i2k\gamma_0~~{\rm with}~~1\leq k\leq  N
\label{nstring}
\end{equation}  
where $\alpha$ is in general complex
with $-\pi/N\leq {\rm Im}\alpha\leq \pi/N.$
We referred to these solutions as exact complete $N$ strings and in
ref.\cite{fm} we numerically presented many examples.
These solutions give a contribution of zero to the energy independent
of $\alpha$ by virtue of the identity
\begin{equation}
\sum_{k=1}^N{1\over \cosh(\alpha+ik2\pi r/N)-\cos r\pi/N}=0
\end{equation}
and thus are responsible for degeneracies in the energy eigenvalue
spectrum.

These exact complete $N$ strings all have the property that in the
equation (\ref{beq}) they  give rise to factors $0/0.$ Therefore
the equation (\ref{beq}) will not be able to determine the parameter
$\alpha$ of the exact complete $N$ string (\ref{nstring}). It is for
this reason that in ref. \cite{fm} we said that Bethe's equation is
incomplete at roots of unity.

However, it is clear that the missing equations can be obtained in
principle by setting
\begin{equation}
\gamma=\gamma_0+\epsilon
\label{eg}
\end{equation}
in (\ref{beq}) and carefully taking the limit $\epsilon\rightarrow 0.$ 
It is the purpose of this paper to carry out this construction which
will complete the specification of the Bethe's roots $v_k.$

For a fixed value of $S^z\geq 0$ any solution $v_k$ 
will have $1\leq k\leq {L\over 2}-S^z.$ If the solution contains $n$
of the complete exact $N$ strings there will be $n_o$ other roots where 
\begin{equation}
n_o={L\over 2}-S^z-nN 
\end{equation}
We denote these roots as $v^{0}_k.$ We will call these
roots ``ordinary'' roots because we will show that they satisfy the
Bethe's equation (where we will let $L$ be even) 
\begin{equation}
\left({\sinh {1\over 2}(v^{0}_j+i\gamma_0)\over \sinh{1\over 2}
(v^0_j-i \gamma_0)}\right)^L=\prod_{l=1\atop l\neq j}^{n_o}
{\sinh{1\over 2}(v^0_j-v^0_l+2i\gamma_0)\over
\sinh{1\over 2}(v^0_j-v^0_l-2i\gamma_0)}.
\label{v0beq}
\end{equation}
This equation for $v^0_k$ does not involve the parameters $\alpha_j$ of
the exact complete $N$ strings and by definition does not have any
ambiguous terms of the form $0/0.$ These ordinary roots 
$v_j^0$ may be real or may have imaginary parts which are
organized into strings (see ref. \cite{fm} for many examples). 
For the purposes of this 
paper this information is not needed and is thus not indicated in the notation.

The simplest case with complete exact $N$ strings is the state which
does not contain any ordinary roots. For a chain  of length $L$ these 
special states have $S^z={L\over 2}-nN$ where $n$ is the number of
exact complete $N$ strings. The $n$ parameters $\alpha_m$ are determined from
the $n$ equations

\begin{eqnarray}
\sum_{k=0}^{N-1}& & \sinh^L{1\over 2}(\alpha_{m} + (2 k
+1)i\gamma_{0})\times \nonumber \\
& &\left(L\coth\frac{1}{2}(\alpha_{m} + (2 k +1)i\gamma_{0}) 
-2 \sum_{l=1\atop l \neq m}^n  \sum_{j=0}^{N-1} 
\coth{1\over 2}(\alpha_{m}-\alpha_{l} +i2j \gamma_{0})\right)= 0
\end{eqnarray}

In general when there are ordinary roots in the state the     $n$
parameters $\alpha_m$ are determined from 
\begin{eqnarray}
 \sum_{k=0}^{N-1}& & \sinh^L{1\over 2}(\alpha_{m} + (2 k +1)i\gamma_{0})
\prod_{l=2k+4}^{N+2k+1}P_{l}(\alpha_{m})\times \nonumber \\
& &\left( L \coth\frac{1}{2}(\alpha_{m} + (2 k +1)i\gamma_{0})  
-2 \sum_{l=1 \atop l \neq m}^n  \sum_{j=0}^{N-1} 
\coth\frac{1}{2}(\alpha_{m}-\alpha_{l} +i2j \gamma_{0})\right.\nonumber \\
&-&\left.\sum_{l=1}^{n_o}( \coth \frac{1}{2}(\alpha_{m}-v_{l}^{0} 
+ 2k i \gamma_{0}) +\coth \frac{1}{2}(\alpha_{m}-v_{l}^{0} 
+ 2(k+1) i \gamma_{0}))\right) = 0
\label{ord1}
\end{eqnarray}
where 
\begin{equation}
P_{k}{(\alpha_{m})} =\prod_{l=1}^{n_o}
\sinh \frac{1}{2}(\alpha_{m}-v_{l}^{0}+ 2 ik \gamma_{0})
\label{pdef}
\end{equation}
and we note the periodicity $P_{k+N}(\alpha_k)=P_k(\alpha_k).$

In the special case $N=2$ the product $\prod_l P_l(\alpha_m)$ is replaced by
one and setting
\begin{equation}
z_{m} = \exp(\alpha_{m}) \hspace{0.5 in} \zeta_{m} = \exp(v^0_{m})
\end{equation}
the equation (\ref{ord1}) reduces to 
\begin{equation}
L (z_{m}^{2}+1)\frac{(z_{m}+i)^{L-2}+(i)^L(z_{m}-i)^{L-2}}{(z_{m}+i)^{L}+
(i)^L(z_{m}-i)^{L}}
-4 \sum_{l=1\atop \neq m}^{n}\frac{z_{m}^{2}+z_{l}^{2}}
{z_{m}^{2}-z_{l}^{2}}
-2\sum_{l=1}^{n_o}\frac{z_{m}^{2}+\zeta_{l}^{2}}{z_{m}^{2}
-\zeta_{l}^{2}}=0.
\end{equation}
It is to be noted that this equation is not the equation which
is obtained by replacing the righthand side of (\ref{beq}) by 
$(-1)^{L/2-|S^z|-1}$ which is what would be obtained if we ``formally''
set $\gamma=\pi/2.$ We also note that the case $N=2$ has been treated 
in ref. \cite{bie} in quite a different context.

These results will be derived in section 2. We will see that in order
to determine the parameters $\alpha_j$ we need to expand (\ref{beq})
to order $\epsilon ^2$ and that the equation (\ref{ord1}) is obtained
as a consistency condition arising from the vanishing of the secular
determinant of a set of homogeneous linear equations.
 
We will discuss the
consequences of our result and the possible relations with the 
evaluation parameters introduced in ref. \cite{fm} in sec. 3.
In particular we present compelling evidence that for $N\geq 3$
the $sl_2$ loop algebra is powerful enough to determine not only 
the structure of the degenerate multiplets but the highest weight 
vectors themselves. On the other hand the highest weight vectors 
are already known from the Bethe's ansatz solution of the XXZ spin chain.
Therefore we conclude that the Bethe's equation (\ref{beq}) is in 
fact contained in
the irreducible representations of $U({\hat sl}_2)$ at level zero.

\section{Derivation of (1.11)}

We begin our derivation by first considering (\ref{beq}) directly at
$\gamma=\gamma_0$ where $v_j$ is an ordinary root $v_j^0$ and split
the product over $l$ on the right hand side into the $n_o$ contributions from 
ordinary roots and the $nN$ contributions from the $N$ strings to find
\begin{eqnarray}
& &\left({\sinh {1\over 2}(v^0_j+i\gamma_0)\over \sinh{1\over 2}
(v^0_j-i \gamma_0)}\right)^L=\nonumber \\
& &\prod_{l=1\atop l\neq j}^{n_o}
{\sinh{1\over 2}(v^0_j-v^0_l+2i\gamma_0)\over\sinh{1\over
2}(v^0_j-v^0_l-2i\gamma_0)}
\prod_{m=1}^{n}\prod_{k=1}^N
{\sinh{1\over 2}(v^0_j-\alpha_m-2i(k-1)\gamma_0)
\over\sinh{1\over 2}(v^0_j-\alpha_m-2i(k+1)\gamma_0)}.
\label{beq2}
\end{eqnarray}
The product over $k$ gives unity independent of $\alpha_m$ and thus
(\ref{beq2}) reduces to (\ref{v0beq}) as desired.

In order to derive (\ref{ord1})
we cannot directly set
$\gamma=\gamma_0$ in (\ref{beq}) and let
$v_j=v_{j,k}=\alpha_j+2ik\gamma_0$ because factors of $0/0$ will occur
on the right hand side. Therefore we first write (\ref{beq})  
\begin{equation}
\sinh^L{1\over 2}(v_{j,k}-i\gamma) q(v_{j,k}+2 i \gamma) + 
\sinh^L{1\over 2}(v_{j,k}+i\gamma) q(v_{j,k}-2 i \gamma) =0
\label{beq3}
\end{equation}
with
\begin{equation}
q(v_{j,k}\pm 2i\gamma) = \prod_{i=1}^{n_o} \sinh
\frac{1}{2}(v_{j,k}-v^0_{i}\pm 2i\gamma)
\prod_{l=1}^{n}\prod_{m=1}^N\sinh{1\over2}(v_{j,k}-v_{l,m}\pm2i\gamma).
\label{qfun}
\end{equation}
We then define for the roots which become exact $N$ strings 
\begin{equation}
v_{j,k}=\alpha_j+2ik\gamma_0+\epsilon
v_{j,k}^{(1)}+\epsilon^{2} v^{(2)}_{j,k}.
\end{equation}
and for all other ordinary roots 
\begin{equation}
v_l=v^0_l+\epsilon v^{(1)}_l.
\end{equation}

We proceed in several steps.

\bigskip

{\bf Expansion of $\sinh^L{1\over 2}(v_{j,k}\pm i\gamma).$}

\bigskip

We expand $\sinh^L{1\over 2}(v_{j,k}\pm i\gamma)$  
to order $\epsilon$ as
\begin{eqnarray}
 \sinh^L{1\over 2}(v_{j,k}\pm i\gamma)
&\sim& \sinh^L{1\over 2}\left(\alpha_j+
2ik\gamma_0+\epsilon v_{j,k}^{(1)}\pm i(\gamma_0+\epsilon)
\right)\nonumber \\
&\sim& \sinh^L{1\over
2}\left[\alpha_j+(2k\pm 1)i\gamma_0\right]\left(
1+\epsilon L{1\over 2}(v_{j,k}^{(1)}\pm i)
\coth{1\over 2}[\alpha_j+(2k\pm 1)i\gamma_0]\right).
\end{eqnarray}

\bigskip

{\bf Expansion of $q(v_{j,k}\pm2i\gamma).$}

\bigskip

To  expand $q(v_{j,k}\pm2i\gamma)$ we 
write
\begin{equation}
q(v_{j,k}\pm2i\gamma)=f(v_{j,k}\pm2i\gamma)\prod_{l=1\atop l\neq
j}^{n}g_l(v_{j,k}\pm2i\gamma)\prod_{l=1}^{n_o}h_l(v_{j,k}\pm2i\gamma)
\end{equation}
with
\begin{eqnarray}
f(v_{j,k}\pm2i\gamma)&=&\prod_{m=1}^N\sinh{1\over 2}(v_{j,k}
-v_{j,m}\pm2i\gamma)\nonumber \\
g_l(v_{j,k}\pm2i\gamma)
&=&\prod_{m=1}^N\sinh{1\over 2}(v_{j,k}-v_{l,m}\pm2i\gamma)~~{\rm
with}~~j\neq l\nonumber \\
h_l(v_{j,k}\pm2i\gamma)&=&\sinh{1\over2}(v_{j,k}-v^0_l\pm2i\gamma)
\end{eqnarray}
and expand $f,~g_l$ and $h_l$ separately.
We find
\begin{eqnarray}
f(v_{j,k}\pm2i\gamma)&=&
\prod_{m=1}^N\sinh{1\over
2}\left(2ik\gamma_0+\epsilon v^{(1)}_{j,k}+\epsilon^2v_{j,k}^{(2)}
-2im\gamma_0-\epsilon
v^{(1)}_{j,m}-\epsilon^2v_{j,m}^{(2)}\pm2i(\gamma_0
+\epsilon)\right)\nonumber \\
&=&\prod_{m=1}^N\left(\sinh[i(k-m\pm 1)\gamma_0]+\right.\nonumber \\
& &\mbox{}\left.+[\epsilon(v_{j,k}^{(1)}-v_{j,m}^{(1)}\pm
2i)+\epsilon^2(v^{(2)}_{j,k}-v^{(2)}_{j,m})]
{1\over 2}\cosh (i(k-m\pm 1)\gamma_0)\right)\nonumber \\
&=&{1\over 2}(-1)^{r(N-k\pm 1)}\left(\prod_{m=1}^{N-1}\sinh mi\gamma_0\right)
\left( \epsilon(v_{j,k}^{(1)}-v_{j,k\pm1}^{(1)}\pm 2i)
+\epsilon^2(v_{j,k}^{(2)}-v^{(2)}_{j,k\pm 1})\right.\nonumber \\
& &\mbox{}\left.+(v_{j,k}^{(1)}-v_{j,k\pm1}^{(1)}\pm2i)
\epsilon^2\sum_{l=1}^{N-1} {1\over 2}
(v_{j,k}^{(1)}-v_{j,k\pm(1+l)}^{(1)}\pm2i)\coth il\gamma_0\right),
\end{eqnarray}

\begin{eqnarray}
g_l(v_{j,k}\pm2i\gamma)&=&\prod_{m=1}^N\sinh{1\over 2}
(v_{j,k}-v_{l,m}\pm2i\gamma)\nonumber\\
&=&\prod_{m=1}^N\left(\sinh{1\over
2}(\alpha_j-\alpha_l+2i(k-m\pm1)\gamma_0
+\epsilon[v_{j,k}^{(1)}-v_{l,m}^{(1)}\pm2i]\right)\nonumber \\
&=&(-1)^{r(N-k\pm1)}\prod_{m=1}^N\sinh{1\over
2}\left(\alpha_j-\alpha_l+2im\gamma_0\right)\times \nonumber\\
& &\mbox{}\left(1+\epsilon\sum_{m=1}^N{1\over 2}
[v_{j,k}^{(1)}-v_{l,m}^{(1)}\pm2i]
\coth {1\over 2}(\alpha_j-\alpha_l+2i(k-m\pm1)\gamma_0)
\right)
\end{eqnarray}
and
\begin{eqnarray}
h_l(v_{j,k}\pm 2i\gamma)
&=&\sinh{1\over2}\left(\alpha_j-v^0_l+2i(k\pm1)\gamma_0
+\epsilon(v^{(1)}_{j,k}-v_l^{(1)}\pm2i)\right)\nonumber\\
&=&\sinh{1\over
2}\left(\alpha_j-v^0_l+2i(k\pm1)\gamma_0\right)\times\nonumber \\
& &\mbox{}\left(1+\epsilon(v_{j,k}^{(1)}-v_l^{(1)}\pm2i){1\over 2}\coth{1\over
2}(\alpha_j-v_l^0+2i(k\pm 1)\gamma_0)\right).
\end{eqnarray}
where we have defined
\begin{equation}
v^{(i)}_{j,N+1}=v^{(i)}_{j,1},~~~v^{(i)}_{j,N  }=v^{(i)}_{j,0} 
~~{\rm with}~i=1,2.
\end{equation}

\bigskip

{\bf Expansion of equation (2.2) to order $\epsilon$}

\bigskip

We now are able to expand (\ref{beq3}). Because $f(v_{j,k}\pm
2i\gamma)$ vanishes as $\gamma\rightarrow \gamma_0$ the leading term
is of order $\epsilon.$ It is convenient to define
\begin{equation}
x_{j,k}=v_{j,k}^{(1)}-v_{j,k+1}^{(1)}+2i~~(-x_{j,k-1}
=v^{(1)}_{j,k}-v^{(1)}_{j,k-1}-2i)
\label{xkdef}
\end{equation}
with
\begin{equation}
\sum_{k=1}^Nx_{j,k}=2iN
\label{norm}
\end{equation}
and also let
\begin{equation}
\phi_k(\alpha_j)=\sinh^L{1\over 2}(\alpha_j+(2k-1)i\gamma_0)
\end{equation}
where we note the periodicity $\phi_{k+N}(\alpha_k)=\phi_k(\alpha_k)$
Then we find that for each value of $j$ the equation (\ref{beq3})
reduces to leading order in $\epsilon$ to a set of $N$ homogeneous
equations for the variables $x_{j,k}$ with $1\leq k \leq N$
\begin{equation}
\phi_{k}(\alpha_j)P_{k+1}(\alpha_j)x_{j,k}
-\phi_{k+1}(\alpha_j)P_{k-1}(\alpha_j)x_{j,k-1}=0
\label{e1set}
\end{equation}
where we recall the definition of $P_k(\alpha_j)$ of (\ref{pdef}).
The determinant of the coefficients of this set of linear equations
vanishes for all values of the still undetermined $\alpha_j$. Thus
equations (\ref{e1set}) are consistent and with the normalization
condition (\ref{norm}) we find $x_{j,k}$ in terms
of $\alpha_j$ as 
\begin{equation}
x_{j,k}={2iN\over K(\alpha_j)}{\phi_{k+1}(\alpha_j)\over
P_{k}(\alpha_j)P_{k+1}(\alpha_j)}
\label{xsol}
\end{equation}
where 
\begin{equation}
K(\alpha_j)=\sum_{k=1}^{N}{\phi_{k+1}(\alpha_j)\over
P_{k}(\alpha_j)P_{k+1}(\alpha_j)}.
\end{equation}

We of course have not yet found  the desired equations to
determine $\alpha_j.$ To do this we need to expand (\ref{beq3}) to one
more order in $\epsilon.$

\bigskip

{\bf Expansion of equation (2.2) to order $\epsilon^2$}

\bigskip

We now define
\begin{equation}
y_{j,k}=v^{(2)}_{j,k}-v^{(2)}_{j,k+1}
\end{equation}
and find
\begin{equation}
\phi_{k}(\alpha_j)P_{k+1}(\alpha_j)y_{j,k}
-\phi_{k+1}(\alpha_j)
P_{k-1}(\alpha_j)y_{j,k-1}=-R_k
\label{e2set}
\end{equation}
where
\begin{equation}
R_k=R_k^{(1)}+ R_k^{(2)}+ R_k^{(3)}+ R_k^{(4)}
\label{rkd}
\end{equation}
with
\begin{eqnarray}
R_k^{(1)}&=&{L\over 4}(v^{(1)}_{j,k}-i)\coth{1\over2}
[\alpha_j+(2k-1)i\gamma_0]\phi_{k}(\alpha_j)P_{k+1}(\alpha_j)x_{j,k}\nonumber
\\
&-&{L\over 4}(v^{(1)}_{j,k}+i)\coth{1\over2}
[\alpha_j+(2k+1)i\gamma_0]\phi_{k+1}(\alpha_j)
P_{k-1}(\alpha_j)x_{j,k-1},
\label{r1}
\end{eqnarray}
\begin{eqnarray}
R_k^{(2)}&=&{1\over4}\sum_{l=1}^{N-1}(v^{(1)}_{j,k}-
v^{(1)}_{j,k+1+l}+2i)\coth il\gamma_0
\phi_{k}(\alpha_j)P_{k+1}(\alpha_j)x_{j,k}\nonumber \\
&-&{1\over 4}\sum_{l=1}^{N-1}(v_{j,k}^{(1)}-v_{j,k-1-l}^{(1)}-2i)
\coth il\gamma_0 \phi_{k+1}(\alpha_j)P_{k-1}(\alpha_j)x_{j,k-1},
\label{r2}
\end{eqnarray}
\begin{eqnarray}
R_k^{(3)}&=&{1\over 4}\sum_{l=1\atop l\neq j}^n
\sum_{m=1}^{N}(v^{(1)}_{j,k}-v^{(1)}_{l,m}+2i)
\coth{1\over 2}(\alpha_j-\alpha_l+2(k-m+1)i\gamma_0)
\phi_k(\alpha_j)P_{k+1}(\alpha_j)x_{j,k}\nonumber \\
&-&{1\over 4}\sum_{l=1\atop l\neq
j}^n\sum_{m=1}^{N}(v^{(1)}_{j,k}-v^{(1)}_{l,m}-2i)
\coth{1\over 2}(\alpha_j-\alpha_l+2(k-m-1)i\gamma_0)
\phi_{k+1}(\alpha_j)P_{k-1}(\alpha_j)x_{j,k-1}\nonumber \\
\label{r3}
\end{eqnarray}
and
\begin{eqnarray}
R_k^{(4)}&=&{1\over 4}\sum_{l=1}^{n_o}(v_{j,k}^{(1)}-v_l^{(1)}+2i)
\coth {1\over 2}(\alpha_j-v^0_{l}+2(k+1)i\gamma_0)
\phi_k(\alpha_j)P_{k+1}(\alpha_j)x_{j,k}\nonumber \\
&-&{1\over 4}\sum_{l=1}^{n_o}(v_{j,k}^{(1)}-v_l^{(1)}-2i)
\coth {1\over 2}(\alpha_j-v^0_{l}+2(k-1)i\gamma_0)
\phi_{k+1}(\alpha_j)P_{k-1}(\alpha_j)x_{j,k-1}.
\label{r4}
\end{eqnarray}

The left hand side of (\ref{e2set}) is identical with the left hand
side of the order $\epsilon$ equation (with $x_{j,k}\rightarrow
y_{j,k}$) and thus the $N\times N$ determinant of the 
coefficients of $y_{j,k}$ vanishes. This implies that the $R_k$ must
satisfy the following constraint
\begin{equation}
\sum_{k=1}^N R_{k}(\alpha_j){P_{k}(\alpha_j)
\over \phi_{k}(\alpha_j)\phi_{k+1}(\alpha_j)}=0.
\label{const}
\end{equation}

\bigskip

{\bf The consistency equation for $\alpha_j$}

\bigskip

It remains to substitute the expression for $R_k$ (\ref{rkd}) and 
$x_{j,k}$ (\ref{xsol}) into (\ref{const}). We consider the four terms
$R^{(i)}_k$ separately.

For $R^{(1)}_k$ we find
\begin{eqnarray}
\sum_{k=1}^N{R^{(1)}(\alpha_j)P_{k}(\alpha_j)\over
\phi_k(\alpha_j)\phi_{k+1}(\alpha_j)}&=&{L\over 4}
\sum_{k=1}^N\left((v^{(1)}_{j,k}-i)\coth {1\over 2}
[\alpha_j+(2k-1)i\gamma_0]{P_k(\alpha_j)P_{k+1}(\alpha_j)
x_{j,k}\over \phi_{k+1}}\right. \nonumber \\
& &\mbox{}\left.-(v^{(1)}_{j,k}+i)\coth {1\over
2}[\alpha_j+(2k+1)i\gamma_0]{P_{k-1}(\alpha_j)P_k(\alpha_j)
x_{j,k-1}\over \phi_k}\right)\\
&=&{NL\over 4}{2i\over K(\alpha_j)}
\sum_{k=1}^N\left( (v^{(1)}_{j,k}-i)\coth{1\over
2}[\alpha_j+(2k-1)i\gamma_0]\right.\nonumber\\
& &\mbox{}\left.-(v^{(1)}_{j,k}+i)
\coth{1\over 2}[\alpha_j+(2k+1)i\gamma_0]\right) \label{r1a}\\
&=&-{iNL\over2 K(\alpha_j)}\sum_{k=1}^N\coth{1\over2}
[\alpha_j+(2k+1)i\gamma_0]x_{j,k}\label{r1b}\\
&=&-{iNL\over 2K(\alpha_j)}\sum_{k=1}^N\coth{1\over
2}[\alpha_j+(2k+1)i\gamma_0]{2iN\phi_{k+1}(\alpha_j)\over K(\alpha_j)
P_{k}(\alpha_j)P_{k+1}(\alpha_j)}\label{r1c}
\end{eqnarray}
where to obtain (\ref{r1b}) we let $k\rightarrow k+1$ in the first
term of (\ref{r1a}) and then use the definition of $x_{j,k}$ and to
obtain (\ref{r1c}) we use (\ref{xsol}). We remark that even though
$v^{(1)}_{j,k}$ appears in $R^{(1)}(\alpha_j)$ of (\ref{r1}) only
the differences $x_{j,k}$ appear in (\ref{r1c}).

For $R^{(2)}_k$ we find
\begin{eqnarray}
\sum_{k=1}^N{R^{(2)}(\alpha_j)P_{k}(\alpha_j)\over
\phi_k(\alpha_j)\phi_{k+1}(\alpha_j)}&=&
{1\over4}
\sum_{k=1}^N\sum_{l=1}^{N-1}\left(v_{j,k}^{(1)}-v_{j,k+1+l}-2i)
\coth il\gamma_0{P_{k}(\alpha_j)P_{k+1}(\alpha_j)\over
\phi_{k+1}(\alpha_j)}x_{j,k}\right.\nonumber\\
& &\mbox{}-\left.(v_{j,k}^{(1)}-v_{j,k-l-1}^{(1)}-2i)\coth il\gamma_0
{P_{k-1}(\alpha_j)P_k(\alpha_j)\over \phi_k(\alpha_j)}x_{j,k-1}\right)
\label{r2a}\\
&=&{iN\over2K(\alpha_j)}\sum_{k=1}^N\sum_{l=1}^{N-1}
(v^{(1)}_{j,k-1-l}-v^{(1)}_{j,k+1+l}+4i)\coth il\gamma_0\label{r2b}\\
&=&0
\label{r2c}
\end{eqnarray}
where to obtain (\ref{r2b}) we have used (\ref{xsol}) and to obtain
(\ref{r2c}) we have let $k\rightarrow k + 2(l+1)$ in 
$v^{(1)}_{j,k-l-1}$ and have used
the antisymmetry of $\coth il\gamma_0$ under $l\rightarrow N-l.$

For $R^{(3)}_k$ we find
\begin{eqnarray}
& &\mbox{}\sum_{k=1}^N{R^{(3)}(\alpha_j)P_{k}(\alpha_j)\over
\phi_k(\alpha_j)\phi_{k+1}(\alpha_j)}\nonumber\\
& &=\mbox{}
{1\over 4}\sum_{k=1}^N
\sum_{l=1\atop l\neq j}^n
\sum_{m=1}^{N}\left((v^{(1)}_{j,k}-v^{(1)}_{l,m}+2i)
\coth{1\over 2}(\alpha_j-\alpha_l+2i(k-m+1)\gamma_0)
{P_{k}(\alpha_j)P_{k+1}(\alpha_j)\over\phi_{k+1}(\alpha_j)}x_{j,k}\right.
\nonumber \\
& &\mbox{}-\left.
(v^{(1)}_{j,k}-v^{(1)}_{l,m}-2i)
\coth{1\over 2}(\alpha_j-\alpha_l+2i(k-m-1)\gamma_0)
{P_{k-1}(\alpha_j)P_{k}(\alpha_j)\over \phi_{k}(\alpha_j)}
x_{j,k-1}\right)\label{r3a} \\
&=&{iN\over2 K(\alpha_j)}
\sum_{k=1}^N \sum_{l=1\atop l\neq j}^n
\sum_{m=1}^{N}\left((v^{(1)}_{j,k}-v^{(1)}_{l,m}+2i)
\coth{1\over 2}(\alpha_j-\alpha_l+2i(k-m+1)\gamma_0)\right.\nonumber \\
& &\mbox{}-\left.
(v^{(1)}_{j,k}-v^{(1)}_{l,m}-2i)
\coth{1\over 2}(\alpha_j-\alpha_l+2i(k-m-1)\gamma_0)\right)\\
&=&-{2N^2\over K(\alpha_j)}\sum_{l=1\atop l\neq
j}^n\sum_{m=1}^N\coth{1\over 2}(\alpha_j-\alpha_l+2im\gamma_0).
\label{r3b}
\end{eqnarray}
We note that even though $v^{(1)}_{j,k}$ appears  in
$R^{(3)}(\alpha_j)$ it has canceled in the final expression (\ref{r3b}). 

Finally for $R^{(4)}_k$ we have
\begin{eqnarray}
\sum_{k=1}^N{R^{(4)}(\alpha_j)P_{k}(\alpha_j)\over
\phi_k(\alpha_j)\phi_{k+1}(\alpha_j)}
&=&{1\over 4}\sum_{k=1}^N\sum_{l=1}^{n_o}\left(
(v_{j,k}^{(1)}-v_l^{(1)}+2i)\coth{1\over2}
(\alpha_j-v_l^{0}+2(k+1)i\gamma_0){P_k(\alpha_j)P_{k+1}(\alpha_j)\over
\phi_{k+1}(\alpha_j)}x_{j,k}\right.\nonumber\\
& &\mbox{}-\left.(v_{j,k}^{(1)}-v^{(1)}_l-2i)\coth{1\over
2}(\alpha_j-v^{0}_l+2(k-1)i\gamma_0){P_{k-1}(\alpha_j)P_{k}(\alpha_j)\over
\phi_k(\alpha_j)}x_{j,k}\right)\label{r4a}\\
&=&{iN\over 2K(\alpha_j)}\sum_{k=1}^N\sum_{l=1}^{n_o}\left(
(v_{j,k}^{(1)}-v_l^{(1)}+2i)\coth{1\over2}
(\alpha_j-v_l^{0}+2(k+1)i\gamma_0)\right.\nonumber\\
& &\mbox{}-\left.(v_{j,k}^{(1)}-v^{(1)}_l-2i)\coth{1\over
2}(\alpha_j-v^{0}_l+2(k-1)i\gamma_0)\right)\label{r4b}\\
&=&{iN\over 2K(\alpha_j)}\sum_{k=1}^N\sum_{l=1}^{n_o}\left(
\coth{1\over2}(\alpha_j-v_l^{0}+2(k+1)i\gamma_0)\right.\nonumber\\
& &\mbox{}\left.+\coth{1\over2}
(\alpha_j-v^{0}_l+2ki\gamma_0)\right)x_{j,k}\label{r4c}\\
&=&{iN\over 2K(\alpha_j)}\sum_{k=1}^N\sum_{l=1}^{n_o}\left(
\coth{1\over2}(\alpha_j-v_l^{0}+2(k+1)i\gamma_0)\right.\nonumber\\
& &\mbox{}\left.+\coth{1\over
2}(\alpha_j-v^{0}_l+2ki\gamma_0)\right){2iN\over
K(\alpha_j)}{\phi_{k+1}(\alpha_j)\over P_k(\alpha_j)P_{k+1}(\alpha_j)}
\label{r4d}
\end{eqnarray}
where we note that $v^{(1)}_l$ (for which we have not derived an
expression) has canceled out in going from (\ref{r4b}) to (\ref{r4c}).

Thus we may use (\ref{r1c}),(\ref{r2c}), (\ref{r3b}) and (\ref{r4d}) in 
(\ref{const}) to obtain
\begin{eqnarray}
& &\mbox{}{1\over K(\alpha_j)}\sum_{k=1}^{N}{\phi_{k+1}(\alpha_j)\over
P_{k}(\alpha_j)P_{k+1}(\alpha_j)}\{L\coth{1\over
2}(\alpha_j+(2k+1)i\gamma_0)\nonumber\\
& &\mbox{}-\sum_{l=1}^{n_o}\left(
\coth{1\over2}(\alpha_j-v_l^{0}+2(k+1)i\gamma_0)
+\coth{1\over 2}(\alpha_j-v^{0}_l+2ki\gamma_0)\right)\}\nonumber \\
& &\mbox{} -2\sum_{l=1\atop l\neq j}^n
\sum_{m=1}^N\coth{1\over 2}(\alpha_j-\alpha_l+2im\gamma_0)=0.
\label{final}
\end{eqnarray}
The result (\ref{ord1}) of the introduction follows immediately. 

\section{Discussion}

The equation (\ref{v0beq}) for the ordinary roots $v^0_l$ and the
equation (\ref{ord1}) for $\alpha_k$ replace the Bethe's equation
(\ref{beq}) at roots of unity where undefined factors of $0/0$
occurred. Thus for the case of $\gamma=r\pi/N$ we have succeeded
for the first time in completely  specifying the parameters
$v_j$ which occur in the Bethe's Ansatz wave function as given by Yang
and Yang \cite{yy} and in the auxiliary matrix $Q(v)$ in Baxter's
\cite{baxc},\cite{baxb} functional equation for the transfer matrix of
the six vertex model. We have also numerically verified for the cases
$N=2-5,~r=1$ that the equations (\ref{v0beq}) and (\ref{ord1}) reproduce
the numerical results previously obtained by other means in ref. \cite{fm}.

There are several other features of our computation which deserve to be
discussed in detail.

First of all even though we have expanded (\ref{beq3}) to order
$\epsilon^2$ and have found an explicit solution (\ref{xsol}) for the
variable $x_{j,k}$ (\ref{xkdef}) which depends on the differences
$v^{(1)}_{j,k}-v^{(1)}_{j,k+1}$ we have not obtained expressions for
the first order corrections $v^{(1)}_{j,k}$ themselves which will in
general contain a constant independent of $k$ which is not present in
$x_{j,k}.$ This constant plays a role in the first order
correction identical to the role which $\alpha_j$ plays in the zeroth
order solution and is determined from the consistency equation needed
for the determination of the $\epsilon^3$ corrections. This pattern of
the consistency equation for order $\epsilon^{n+2}$ being needed to
completely specify the roots to order $\epsilon^n$ is a general
feature for all orders in $\epsilon.$

Secondly we point out that the parameters $\alpha$ introduced in
(\ref{nstring}) as part of the specification of the complete exact $N$
string  do not have to  be real and that if $\alpha$ is complex then
we see from taking the complex conjugate of (\ref{final}) that
$\alpha^*$ will also be a a solution (because all ordinary roots
appear in complex conjugate pairs). These complex values of $\alpha$ are a new
feature of the solution of Bethe's equation which were first seen in
our \cite{fm} previous numerical treatment of the problem where many
examples were exhibited. 
 
Thirdly we acknowledge  that to completely solve the eigenvalue
degeneracy problem we need
to be able to classify and count 
all solutions of (\ref{v0beq})
and (\ref{ord1}).
We studied this degeneracy in ref. \cite{dfm} and \cite{fm} in terms
of the finite dimensional representations of the $sl_2$ loop algebra
and we remarked that from the theory of affine Lie
algebras\cite{cp} it is known that these representations are specified
by what are called ``evaluation'' parameters and that these parameters
are roots of the  Drinfeld polynomial. However these evaluation
parameters are not the same as the roots $\alpha_j$ studied in this
paper and we note in particular that for a multiplet containing $2^m$
degenerate eigenvalues there are only $m$ evaluation parameters even
though there are $2^m$ solutions for $\alpha_j$ all of which have the
same $n_o$ ordinary roots. Thus it would seem
that there is a sense in which there is more information in the
$\alpha_j$ than what is needed for the solution of the degeneracy
problem for the eigenvalues of the Hamiltonian (and the transfer
matrix).

To explore further  the nature of the information contained 
in the evaluation parameters we have explicitly computed them for a chain
of $L=12$ and $N=3$ and have found that the evaluation parameters of each 
multiplet are different. This means that
there are no isomorphisms
between multiplets and thus we reach the striking conclusion  for
this case (and we believe for all cases with $N\geq 3$) that the $sl_2$ 
loop algebra is not only a symmetry algebra which classifies states 
into degenerate multiplets but that it is sufficiently powerful
to determine all the highest weight vectors as well. Therefore the complete
decomposition of the representation of the loop $sl_2$ algebra
into finite dimensional irreducible representations must produce as highest
weight vectors the same vectors which are determined by the Bethe's 
equation (\ref{beq}) when used in the Bethe form of 
the wave function \cite{yy}. This relation between finite dimensional
irreducible representations of the loop algebra of $sl_2$ and 
Bethe's ansatz has not been previously noticed.

Finally we mention that unlike the previous discussion \cite{dfm} 
of the degeneracies of the XXZ Hamiltonian at roots of unity in terms
of the $sl_2$ loop algebra for which the generalization to the XYZ  
model is unknown, the considerations of this present paper have a very
natural extension to the XYZ model. Details of this extension will be
published elsewhere.

\newpage

\centerline{\large\bf Acknowledgments}

We are greatly indebted to 
M. Jimbo, M. Kashiwara, and T. Miwa for many illuminating discussions 
about quantum groups.
This work is supported in part by the National Science Foundation under
Grant No. DMR-0073058.

\end{document}